\documentclass[prb,preprint,superscriptaddress,amsmath,amssymb]{revtex4-1}

\usepackage{graphicx, nicefrac}
\usepackage{array}
\usepackage{amsmath,amsfonts,amssymb}
\usepackage[ansinew]{inputenc}
\usepackage{color, soul}
\usepackage{calc}

\newcommand{\Figref}[1]{Fig.~\ref{#1}}

\newcommand{\ignore}[1]{}

\begin{document}
\title{Tuning single-molecule fluorescence by atomic-scale control of the local environment}

\author{Thiago G. L. Brito}
\affiliation{Max Planck Institute for Solid State Research, Stuttgart, 70569, Germany}

\author{Daniel Arribas}
\affiliation{Max Planck Institute for Solid State Research, Stuttgart, 70569, Germany}

\author{Sofia Canola}
\affiliation{Institute of Physics, Czech Academy of Sciences, Cukrovarnick\'{a} 10, Prague, 16200, Czech Republic}

\author{Klaus Kuhnke}
\affiliation{Max Planck Institute for Solid State Research, Stuttgart, 70569, Germany}

\author{Tom{\'a}\v{s} Neuman}
\affiliation{Institute of Physics, Czech Academy of Sciences, Cukrovarnick\'{a} 10, Prague, 16200, Czech Republic}

\author{Anna Ros\l awska}
\email{a.roslawska@fkf.mpg.de}
\affiliation{Max Planck Institute for Solid State Research, Stuttgart, 70569, Germany}

\begin{abstract}

Molecules that absorb and emit light play a central role in microscopy, light-emitting devices and photosynthesis. Their fluorescence arises from well-defined radiative transitions that are governed by the electronic states and their coupling to the nuclear motion that are influenced by the local environment. Yet the effect of controlled atomic-scale variations in the emitter surroundings remains unexplored. Here, we use scanning tunneling microscopy combined with optical spectroscopy to investigate the optical response of a single phthalocyanine to the change in the position of a nearby molecule, controlled with precision better than 100 pm. Upon decreasing the intermolecular distance, the molecular emission energy redshifts and its line profile evolves. Supported by theoretical calculations, we disentangle the electronic and nuclear contributions to the changes in fluorescence. We find that the redshift originates from the interaction between the excitations of the two molecules, while the lineshape changes reflect modifications of the molecular rotational degree of freedom and non-equilibrium dynamics. We extend this control to larger assemblies, where one molecule tunes the energies of two chromophores, mimicking the environmental tuning in photosynthetic systems. Our study provides atomic-scale insight into how the local environment affects the optical properties of molecular systems.

\end{abstract}

\date{\today}

\maketitle

The fluorescence of a light-emitting molecule originates from well-defined transitions determined by its electronic structure and the coupling to the motion of its nuclei. Besides their intrinsic characteristics, however, the properties of molecular emitters are largely defined by their local environment, including nearby molecules, charges, host matrices, fields, solvents, and interfaces\cite{Orrit1990, Ambrose1991, Kulzer1997, Hettich2002, Colautti2020, Smit2023, Muh2010}. The environment tunes the properties of the emitter at the atomic scale, leading to a series of detectable features in the emission spectra. For example, the molecular excitations can be shifted directly by electronic effects, while modifications of the molecular adsorption geometry and environmental hindrance of nuclear motion can alter the observed spectral shape. These phenomena give rise to dynamical peak broadening due to fast processes causing dephasing or decay, to static shifts of emission lines, and to slow spectral diffusion\cite{Ambrose1991}, which average over time and molecular ensembles into inhomogeneously broadened spectra\cite{Moerner1989}. Identifying the microscopic mechanisms, and thus the dominant sources of dephasing or energy shifts, is a necessary step towards the development of, for example, spectrally stable emitters featuring narrow dephasing-free emission lines\cite{Toninelli2021}, designing artificial light-harvesting complexes\cite{Croce2014, Scholes2011}, or interpreting the contrast in super-resolution microscopy\cite{Bongiovanni2016}. However, systematically addressing processes affecting the fluorophores at the single-molecule level requires both direct access to and control over the environment of the emitter. Such capabilities are beyond the limits of far-field fluorescence or super-resolved techniques\cite{Schermelleh2019}.

Scanning tunneling microscopy (STM) combined with optical spectroscopy provides an ideal toolbox to overcome these limitations.  The tip-based approach allows both manipulation of the local environment of a single molecule\cite{Crommie1993, Zhang2016, Cortes-delRio2020, Leisegang2021} and probing its optical properties with atomic precision owing to the localization and enhancement of the optical electromagnetic field at the tip apex \cite{Roslawska2026}. The molecule can be excited electrically in STM-induced luminescence (STML)\cite{Zhang2016, Imada2016, Doppagne2017, Roslawska2026, Cao2021, Kong2022, Zheng2026, Rai2026, Luo2021, Hung2021, Dolezal2020}, or optically, leading to tip-enhanced photoluminescence (TEPL)\cite{Yang2020, Roslawska2024, Kaiser2024a} and tip-enhanced Raman scattering (TERS) \cite{Zhang2013, Li2021c,deCamposFerreira2026,Liu2020} signals. STML and TEPL further permit probing spectral shifts and broadening due to the fields induced by the tip or point charges\cite{Imada2021, Kuhnke2017a, Arrieta2026, Roslawska2022, Vasilev2022, Sagwal2026, Yang2020, Zhang2017}. Moreover, STML enables the molecule to be excited indirectly via resonant energy transfer (RET)\cite{Imada2016, Cao2021, Kong2022}, which can be used to observe fluorescence of molecules that do not permit efficient direct electrical pumping\cite{Vasilev2022, Hung2024, Li2026}.

\begin{figure}[!ht]
    \includegraphics[width = 0.5\linewidth]{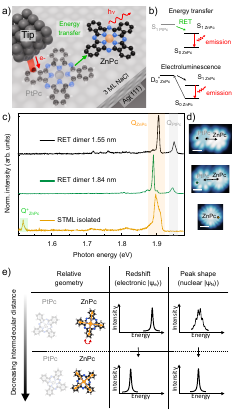}
    \caption{\textbf{Probing the influence of the environment on the optical properties of a single molecule.} a) Schematics of the experiment. ZnPc (the molecule of interest) interacts with a nearby PtPc (grayed out). We excite the fluorescence of ZnPc either directly (not shown) or indirectly via energy transfer from PtPc. b) ZnPc light emission mechanisms. Top: energy transfer; note that the left state is the PtPc $S_1$ state. Bottom: direct STML; all states are ZnPc states. c) Optical spectra recorded at positions indicated in d), $I =$ 300 pA (80 pA for 1.84 nm), $V =$ -2.6 V, $t$ = 120 s (60 s for 1.84 nm). d) STM images, $I =$ 2 pA, $V =$ -2.6 V, scale bars are 1 nm. e) Modifications of the relative position of the two molecules lead to redshifts and peak shape changes of the ZnPc emission. This is due to the changes in the electronic structure and coupling to the nuclear motion represented by the electronic and nuclear parts of the wavefunction, respectively. The red arrow indicates the shuttling of ZnPc.}
    \label{Fig1}
\end{figure}

Here we apply STML to probe a model system consisting of a zinc-phthalocyanine (ZnPc) molecule placed in the vicinity of a platinum-phthalocyanine (PtPc) molecule that acts as a model environment tuning the optical properties of the ZnPc (see \Figref{Fig1}a). We control the interaction strength between the two molecules by manipulating the distance between them with a precision better than 100 pm and measure the fluorescence of ZnPc. As the distance between the molecules in the ZnPc-PtPc dimers is reduced, the ZnPc emission redshifts and its lineshape changes from a spectrum featuring prominent side peaks to a nearly single narrower peak. The spectrum thus loses rather than gains fine structure as the molecules approach. Aided by theoretical models, we identify configuration-dependent evolution of the electronic coupling between the two molecules, modification of the nuclear potential energy landscape, and signatures of dissipation related to the nuclear motion, as summarized in \Figref{Fig1}e. Finally, we apply these mechanisms to tune the transition energy of two ZnPc molecules in a larger structure, similar to the environmental tuning of energies in photosynthetic systems.

To acquire the data, we use a low-temperature (4 K), ultrahigh-vacuum STM with optical access. As illustrated in \Figref{Fig1}a, the molecules adsorb on a thin insulating layer (3 monolayers, ML) of NaCl, which decouples them from the supporting metal and avoids substrate-mediated quenching of the molecular excited states (see Supplementary Information S1 for more details on the set-up). Even though this decoupling preserves the excited states, resolving the changes that PtPc induces in the fluorescence of ZnPc requires minimizing two tip-mediated effects on the emission spectra. First, the interaction between the STM tip and the molecular excited state leads to a shift of the emission peak (Stark and photonic Lamb effects) and peak broadening (Purcell effect) \cite{Roslawska2022, Yang2020, Imada2021, Kuhnke2017a, Zhang2017}, all of which are maximized when the tip is located directly on top of the molecule. Second, exciting the molecule directly by placing the tip on top of it and running a current through the junction yields a broad electroluminescence peak at $\sim$ 1.90 eV (bottom plot in \Figref{Fig1}c), assigned to the radiative decay of its lowest excited state ($Q_{ZnPc}$ transition)\cite{Zhang2016, Doppagne2017}. This complex spectral shape originates from the transient charging of the molecule\cite{Zhang2016} and precludes clear identification of the fundamental emission energy necessary to quantitatively evaluate its shifts. To avoid these issues, we move the tip away from ZnPc and excite the molecule remotely via a RET process from the nearby PtPc. In this process PtPc acts as a donor and ZnPc as an acceptor chromophore \cite{Imada2016, Cao2021, Kong2022}. A detailed description of the mechanisms driving the molecule to the excited state is provided in Supplementary Information S2.

In \Figref{Fig1}c, we present optical spectra recorded while exciting the ZnPc via RET (top and middle plots) in the molecular configurations shown in \Figref{Fig1}d. While these arrangements are defined by both the center-to-center distance and the relative angle, we refer to them for clarity by the distance only, here 1.84 nm and 1.55 nm. Besides the narrow ZnPc emission at 1.90 eV, we also observe the fluorescence of the donor (labeled as $Q_{PtPc}$) at 1.95 eV\cite{Grewal2025a}. Already these broad-range spectra reveal a clear redshift of the emission when PtPc is closer to ZnPc.

\begin{figure}[!ht]
    \includegraphics[width = \linewidth]{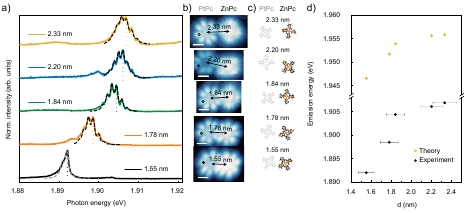}
    \caption{\textbf{Controlling single-molecule fluorescence via the local environment.} a) High-resolution optical spectra of ZnPc recorded while exciting the molecule via RET from PtPc. The distances indicated in the plots refer to the orientations in b) and c). The dashed lines are fits to the model discussed in the text, and the dotted lines indicate the energy of the fundamental $E_{00}$ transition. The spectra are recorded at the positions marked in b). Parameters: $V$ = -2.6 V, $I$ = 300 pA, $t$ = 900 s (2.33 nm), $t$ = 300 s (2.20 nm, 1.55 nm), $t$ = 120 s (1.84 nm), $t$ = 60 s (1.78 nm). b) STM images of the studied dimers, $V$ = -2.6 V, $I$ = 2 pA, scale bars are 1 nm. c) Molecular orientations of the dimers. d) Comparison between the experimental values of $E_{00}$ extracted from a) and the quantum chemistry calculations. The error bar reflects the experimental uncertainty (5 \%) in evaluating the distances.}
    \label{Fig2}
\end{figure}

High-resolution emission spectra of ZnPc as a function of the center-to-center distance ($d$) between the ZnPc and the perturbing PtPc resolve how the ZnPc emission changes in different configurations (see \Figref{Fig2}a). We find that the spectra progressively shift towards lower energy when the PtPc lies closer to the ZnPc. Except for the smallest distance of 1.55 nm, the spectra feature a fine structure characterized by a manifold of equally spaced peaks. Such features have previously been identified in the doublet (trion, $D_1^+ \rightarrow D_0^+$ transition) emission of ZnPc\cite{Dolezal2022} and assigned to the coupling between the exciton and the librations (hindered rotations) of the molecule. This motion is a torsional oscillation around an equilibrium angle without net translation of the center of mass (located on top of a Cl$^-$ ion for ZnPc on NaCl \cite{Dolezal2022}). Here, by exciting the ZnPc remotely, we bypass its transient charging and the complex emission characteristics, and find the librational features also in the emission of the neutral exciton ($S_1 \rightarrow S_0$ transition) of ZnPc. This observation is not limited to the ZnPc-PtPc system, and we find similar features also in the direct TEPL/TERS spectra of an isolated ZnPc molecule and in ZnPc-H$_2$Pc dimers (see Supplementary Information S3 and S4, respectively). 

Such spectra can be fitted using a quantum harmonic torsional oscillator model\cite{Dolezal2022,Kogler2026} based on the Franck-Condon principle, in which the coupling stems from slightly shifted minima of the rotational potentials (i.e., a shift of the adsorption angle of the molecule with respect to the substrate). This shift leads to a non-zero wavefunction overlap between $S_1$ and $S_0$ librational states with different quantum numbers. We first use this model to fit the spectra and disentangle the librational contributions from the fundamental zero-phonon line $E_{00}$ (marked by dotted lines in \Figref{Fig2}a), thereby enabling an evaluation of the emission line shifts. The dashed lines in \Figref{Fig2}a show the resulting fits overlaid on the data. The evolution of the remaining fitting parameters reveals the physical mechanisms captured by this model, and we return to it later in the text (see also Supplementary Information S5).

The extracted $E_{00}$ energy for all studied distances, 2.33 nm $\ge d \ge$ 1.55 nm is shown in \Figref{Fig2}d. The absolute shift of the $E_{00}$ line recorded in this range of distances amounts to 15 meV (see $E_{00}$ values in Supplementary Table II) and demonstrates how the emission energy of a single molecule can be controllably tuned by its local environment with picometer-scale precision. We also observe a redshift in the direct STML measurement of ZnPc (see Supplementary Information S6). To identify the origin of this experimental trend, we perform time-dependent density-functional theory (TDDFT) calculations for the molecular dimers (see Supplementary Information S7 for more details), taking into account the distance and orientation of the ZnPc with respect to the perturbing PtPc (\Figref{Fig2}c). The calculations reveal that the excitations of the molecular monomers hybridize weakly, resulting in dimer excited states localized dominantly on a single monomer only (see Supplementary Information S7 for more details). The emission energies corresponding to the ZnPc-localized excitation are plotted in \Figref{Fig2}d and show excellent agreement with the experimental trend.

The interpretation of the fluorescence experiments and their correlation with the STM images of the PtPc-ZnPc dimers allows the identification of two qualitatively different interaction regimes observed at the extremes of studied $d$ range. First, we observe a weaker interaction regime, exemplified by the representative case of the dimer with an intermolecular distance of $d = 2.33$ nm (upper panel in \Figref{Fig2}b). It is characterized by a fine spectral structure with up to 7 peaks, in which several quanta of the librational modes in $S_{1}$ and $S_{0}$ couple to the emission (upper row of \Figref{Fig1}e), and a distinct appearance in the STM images. At high negative bias voltages, ZnPc appears as a 16-lobe feature as a result of its rapid \textit{shuttling}\cite{Dolezal2022, Peller2020, Patera2019a, Dolezal2019, Hung2023, Hung2021} between two equivalent adsorption angles $\phi \approx$ $\pm$ 11.5$^{\circ}$ with respect to the [011] direction of the NaCl layer, as evaluated from the experiments. Second, we identify a stronger interaction regime, exemplified by the dimer with an intermolecular distance of $d = 1.55$ nm (lower panel in \Figref{Fig2}b). It is characterized by a lower number of libron peaks (one peak with two low-energy shoulders) associated with a sharper fluorescence peak as a result of the coupling to a lower number of quanta of the libration mode (lower row of \Figref{Fig1}e). It also presents a considerable redshift of about 15 meV with respect to the $d = 2.33$ nm case. In STM images, the ZnPc molecule appears \textit{pinned} in one of its two possible configurations and appears instead as an 8-lobe shape similar to PtPc. Notably, other pinned configurations in ZnPc multimers can also result in a sharpening of their emission peaks and coherent coupling\cite{Zhang2016, Doppagne2017}.

The transition between the weaker and stronger interaction regimes is gradual. Qualitatively, the fine structure of the fluorescence peak undergoes a subtle, progressive decrease in the relative intensities of the high-energy components (at energies above $E_{00}$, also termed hot luminescence\cite{Roslawska2026}) without abrupt changes in its general shape, indicating that a progressively lower number of libron quanta are involved (a quantitative analysis is presented below). This suggests that the interaction with PtPc acts as a perturbation that favors the dissipation of energy from the libron modes in ZnPc without substantially affecting the potential landscape around their equilibrium angles, at which the exciton recombination takes place. In STM images, we find that ZnPc does not shuttle at a separation of $d= 1.78$ nm, which we tentatively attribute to further restriction of the rotational degree of freedom of the ZnPc nuclei, resulting in a \textit{hindered} configuration (see also Supplementary Information S8 for further discussion). Likewise, adsorption next to step edges can result in a similar configuration\cite{Dolezal2022, Hung2021} (see also Supplementary Information S9).

\begin{figure}[!ht]
    \includegraphics[width = \linewidth]{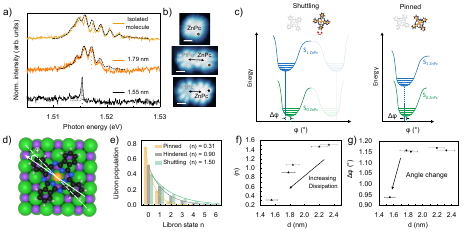}
    \caption{\textbf{Analysis of the nuclear effects modifying the fluorescence spectra.}  a) STML spectra of the $D_1^+$ emission of ZnPc excited directly at positions marked in b). The dashed lines are fits to the model. Parameters: $V$ = -3.0 V and $I$ = 300 pA. $t$ = 1500 s (top), $t$ = 5400 s (middle), $t$ = 600 s (bottom). b) STM images of the studied ZnPc monomer and ZnPc-PtPc dimers, $V$ = -2.6 V, $I$ = 2 pA.  All scale bars are 1 nm. c) Top: molecular structures for the shuttling and pinned ZnPc molecules, the red arrow indicates the shuttling motion. Bottom: Schematics of the $S_0$ and $S_1$ potential energy landscapes for both configurations. The populated libronic states are located close to the minima of the potential wells. The mismatch in the angles between the minima of $S_0$ and $S_1$ potentials leads to the observable libronic features in the optical spectra. For completeness, in the shuttling configuration, we also represent the second potential well corresponding to the other ZnPc adsorption configuration. Note that the transition energy shifts and changes in $\Delta \phi$ are exaggerated for illustrative purposes. d) The adsorption configuration of ZnPc on NaCl. e) Populations of the bottom libronic states in the shuttling ($d=$ 2.33 nm), hindered ($d=$ 1.78 nm), and pinned ($d=$ 1.55 nm) configurations. The solid lines are guides to the eye. f,g). $\langle n\rangle$ (f) and $\Delta \phi$ (g) as a function of $d$.}
    \label{Fig3}
\end{figure}

A much more prominent fingerprint of the environmental effects on the high-energy libron components is present in the evolution of the libronic fine structure of the doublet emission of cationic ZnPc ($Q^+_{ZnPc}$ at 1.52 eV in Fig. 1c)\cite{Doppagne2018, Kaiser2025, Dolezal2021a}. This is represented in \Figref{Fig3}a, which shows $D_1^+ \rightarrow D_0^+$ emission spectra for an isolated shuttling ZnPc (\Figref{Fig3}a top), a hindered ZnPc in a dimer (\Figref{Fig3}a center), and a pinned ZnPc in a dimer (\Figref{Fig3}a bottom). \Figref{Fig3}b displays the corresponding STM images. In contrast to the data presented in \Figref{Fig2}a, here we record the spectra directly on ZnPc, as driving the molecule to the $D_1^+$ state requires two charge exchanges with the tip, a process in which the triplet state acts as a relay \cite{Kaiser2025} (see Supplementary Information S2). Since the molecule has to be transiently charged and excited simultaneously, such an emission has not been observed in STML via RET. Additionally, in the case of cationic ZnPc, the spectra are much sharper than those obtained via directly excited emission of the $S_1 \rightarrow S_0$ transition, allowing for resolution of its libronic fine structure even when the molecule is isolated. As for the neutral emission reported in \Figref{Fig2}, we observe a redshift of the main emission line (in a total range of 2 meV, lower than the 15 meV observed in the $S_1 \rightarrow S_0$ transition) and a significant decrease in intensity of the high-energy components (at energies above $E_{00}$) upon interaction with a neighboring PtPc molecule. In the stronger interaction regime, we observe only one sharp (0.3 meV full-width at half maximum) peak without apparent libronic fine structure. We fit the spectra using the model described above and obtain excellent agreement with the experimental data. Compared to the neutral emission, we find that the spacing observed between the peaks in the experimental spectra changes from the range 1.1-1.4 meV in the neutral transition to around 1.4-1.8 meV, reflecting an increase in the stiffness of the rotational potentials in agreement with calculations (see Supplementary Information S5)\cite{Dolezal2022}.

Having probed how the different configurations of the ZnPc and its environment affect the spectral features of the ZnPc emission, we now turn to a detailed analysis of the physical mechanisms underlying these effects using the harmonic torsional oscillator model. \Figref{Fig3}c illustrates the Franck-Condon scheme for the shuttling and pinned scenarios. The potential surfaces (solid lines) of the ground and excited states can be approximated as parabolas described by a set of fitting parameters. Besides $E_{00}$, these include the stiffness of the harmonic molecule-surface interaction potential ($k_0$ for the ground state and $k_1$ for the excited state), the expected number of excited libron quanta ($\langle n\rangle$), the angular displacement ($\Delta \phi$) between the minima of the parabolas, and the emission line broadening $\sigma$. Small $k$ results in an opening of the parabola, while a large value leads to steeper characteristics (see Fig. S4). The adsorption angle of ZnPc on NaCl ($\phi$) is illustrated in \Figref{Fig3}d. We fit our measurements with this model by restricting the number of free parameters to the minimum. The stiffnesses $k_0$ and $k_1$ are guided by first-principles calculations\cite{Dolezal2022}, and the main parameters controlling the shape of the spectra are $\langle n\rangle$ and $\Delta \phi$. $E_{00}$ only shifts the position on the energy axis, and $\sigma$ controls the width of each individual feature. \Figref{Fig3}f and g show the evolutions of $\langle n\rangle$ and $\Delta \phi$ as a function of $d$ for the $S_1 \rightarrow S_0$ transition. We find that in the weaker interaction regime the change in the expected number of excited librons drives the evolution of the spectral shapes that mostly differ by the distribution of weight between the peaks forming the full comb structure. We remark that the number of excitations is relatively low (see \Figref{Fig3}e for the population distribution), so that the libron wavefunctions are localized near the minima of the potentials, ensuring that the parabolic approximation is valid. While for $d >$ 1.55 nm $\Delta \phi$ varies only minimally, we find that upon the transition to the stronger interaction regime, where the ZnPc is in a pinned configuration, $\Delta \phi$ changes from 1.15$^{\circ}$ to 0.94$^{\circ}$, which, together with a decrease in $\langle n\rangle$, further reduces the number of observed libronic peaks. We find a similar trend for the $D_1^+ \rightarrow D_0^+$ transition (\Figref{Fig3}a, see also Table III in the Supplementary Information). Based on this analysis we draw two main conclusions regarding the influence of the environment, here a perturbing PtPc molecule, on the nuclear component of the fluorescence spectra. First, the reduction of the average number of quanta of libron excitation in $S_{1}$ and $D_1^+$ upon approaching the PtPc indicates that this neighboring molecule assists in the energy dissipation from the ZnPc and changes the non-equilibrium dynamics within the system. Tentatively, we attribute the dissipation to the coupling to the librations of PtPc. Second, the abrupt transition in the $\Delta \phi$ upon pinning the molecule stems from an interaction with PtPc strong enough to significantly perturb the rotational potential energy landscape describing the ZnPc, which we interpret as an effect related to the steric hindrance (due to close proximity between the hydrogen atoms of the two molecules, see more in Supplementary Information S10) that also restricts the shuttling motion of ZnPc. Both mechanisms reduce the observed number of features, and the spectrum therefore loses rather than gains fine structure as the molecules approach.

\begin{figure}[!ht]
    \includegraphics[width = 0.5\linewidth]{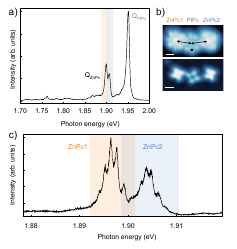}
    \caption{\textbf{Tuning the emission energy in larger assemblies.} a) STML wide energy range spectrum recorded on the PtPc (position marked in b)) in a ZnPc-PtPc-ZnPc trimer. b) STM images of the ZnPc-PtPc-ZnPc trimer. Top: $V$ = -2.6 V, $I$ = 2 pA, bottom $V$ = -1.2 V, $I$ = 2 pA. Scale bars are 1 nm. c) High-resolution STML spectrum recorded at the position indicated in b). Parameters for the STML spectra: $V$ = -2.6 V, $I$ = 300 pA and $t$ = 60 s. }
    \label{Fig4}
\end{figure}

As a final step, we tune the emission energies of chemically identical emitters in a larger assembly, in analogy to photosynthetic systems in which the protein and pigment environment tunes chlorophyll energies to direct energy transfer\cite{Muh2010}. We investigate a molecular trimer consisting of two ZnPc molecules that are separated by a PtPc (\Figref{Fig4}). \Figref{Fig4}b shows the STM image of the structure in which one ZnPc (labeled ZnPc1) is located at a distance of 1.77 nm from the PtPc (hindered configuration) whereas the other ZnPc (ZnPc2) is slightly further away at 2.01 nm (shuttling configuration). The detailed arrangement of the trimer and the relative molecular orientations become clear in the in-gap STM image (\Figref{Fig4}b). \Figref{Fig4}a shows a low-resolution spectrum recorded on the PtPc. Besides the emission of PtPc ($Q_{PtPc}$) followed by its vibrational progression, it exhibits a double-peak feature around 1.9 eV. Upon closer examination of a high-resolution spectrum (\Figref{Fig4}c), we find two emission components featuring the libronic progression with emission energies centered at 1.905 eV for $d=$ 2.01 nm and 1.899 eV for $d=$ 1.77 nm. This is consistent with the trend shown in \Figref{Fig2} and allows us to identify these emission lines as originating from ZnPc1 (lower energy and closer to PtPc) and ZnPc2 (higher energy and further away from PtPc). Furthermore, we find that ZnPc2 can transfer energy to ZnPc1, a process in which the larger band-gap species (PtPc) acts as a passive molecule\cite{Cao2021} (see Supplementary Information S11). 

In summary, our work demonstrates that the fluorescence of a single ZnPc molecule can be tuned by engineering the local atomic-scale environment with picometer-scale precision. We find that the presence of another molecule nearby affects the emission energy and fine structure, reflecting contributions arising from the electronic coupling and from the nuclear motion (\Figref{Fig1}e), such as dissipation and modifications of the potential energy landscape. Resolving these effects at the level of model molecular dimers and trimers, as demonstrated here, offers a route to identifying the microscopic mechanisms that govern the optical properties of complex molecular and solid-state systems.

\section*{Acknowledgments}

We thank J. Friese, K. Kern, and I. Padniuk for fruitful discussions, and W. Stiepany, and M. Memmler for technical support. T. G. L. B. and A. R. acknowledge funding from the Emmy Noether Programme of the Deutsche Forschungsgemeinschaft (DFG, German Research Foundation; Grant No. 534367924). T. N. and S. C. acknowledge support from the Lumina Quaeruntur fellowship of the Czech Academy of Sciences. 

\bibliographystyle{naturemag}
\bibliography{references}

\end{document}